\documentclass[floats,prb,aps,groupedaddress]{revtex4}
\usepackage{epsfig}

\begin{document}
\title{Electromigration theory unified}

\author{A. Lodder}

\affiliation{Faculty of Sciences/Natuurkunde en Sterrenkunde, Vrije Universiteit De
Boelelaan 1081, \\ 1081 HV Amsterdam, The Netherlands}

\date{\today}

\begin{abstract}
\normalsize{
The starting formula of Bosvieux and Friedel \cite{quantum} for the force
on an ion in a metal
due to an applied voltage is shown to lead to the same description
as the linear-response approach used in the field since its
introduction by Kumar and Sorbello \cite{kumar}. By this
electromigration theory has become a unified theory. It follows after
accounting for a treacherous trap term, which at first sight
seems to be zero. Up to now
Bosvieux and Friedel claimed to predict a completely screened
direct force, which means that only a wind force would be operative.
In addition the amount of screening has been calculated up to
second order in the potential of the migrating impurity,
using a finite temperature version of the screening term derived by Sham \cite{sham}.
For a proton in a metal modeled as a jellium the screening
appears to be about 15\%, which is neither negligible nor
reconcilable with the old full-screening point of view.
}
\end{abstract}

\pacs ~~~~PACS numbers are 66.30.Qa and 72.10.Bg

\maketitle

\section*{Introduction}

Electromigration is the motion of an ion in a metal under the influence of
an applied voltage. Depending on the metal and the ion
its motion is either to the anode or to the cathode. The total driving force
on such an ion is known to be
the result of two contributions, a direct force and a wind force
\begin{equation}
\label{FZster}
{\bf F}= {\bf F}_{direct}+{\bf F}_{wind}=
(Z_{direct}+Z_{wind})e{\bf E}=Z^* e{\bf E},
\end{equation}
in which $Z^*$ is called the effective valence, which is a
measurable quantity.
The direct force is due to the direct action of the applied field
${\bf E}$ on the charge of the ion. The wind force comes
from the scattering of the current carrying electrons off the ion.\cite{sor97}

Until 1962 there was a common belief that $Z_{direct}$ was equal
to the bare valence $Z_{i}$ of the ion. At most
a small deviation from that value could arise from the electrons
in the metallic environment of the ion. In 1962 Bosvieux and Friedel \cite{quantum}
predicted a complete cancellation of the direct force due to screening
effects, so that only the wind force would remain. By that
prediction a controversy was born. It was not easy to decide matters
by a measurement, and a satisfactory theoretical answer was lacking
as well. 

In 1975 Kumar and Sorbello \cite{kumar} published an exact
linear-response expression for the driving force. After that
it was considered as being just a matter of a careful evaluation
of that expression in order to settle the problem. 
It took quite some time to do so though. Finally Sorbello \cite{sor85},
starting from the complicated treatment given by
Rimbey {\it et al.}\cite{rimbey}, predicted a screening of
10-30\%, depending on the potential used for the ion.
Support of this result has been given by the present author,
starting from a much simpler description.\cite{lodderKrakow} However,
in the meantime new support was given for the full-screening point of view.
\cite{turban}

Interestingly, the original paper by Bosvieux and Friedel (BF) is referred to
as being controversial on the one hand, while on the other hand their starting formula
is referred to as the first quantum mechanical
equation for the wind force, which is seen as a pioneering contribution 
to the field. Therefore we undertook an new evaluation of their starting formula,
following the authors as closely as possible, up to a point where
we came across a trap. By treating this trap properly the final evaluation ends up
at the standard linear-response expression for the driving force.
As far as the author knows Sorbello was the first who showed that
such traps occur in the theory of electromigration.\cite{sorbelloObj}

The screening itself has been calculated
by Sorbello\cite{sor85}, who used one type of expression for it. Therefore
we undertook the evaluation of Sham's second order expression
written in a new form, the more so as a more recent result for the screening,
ranging from 0 to 100\%, is rather inconclusive.\cite{ishida}
Sham's expression has never been
evaluated numerically, and in his paper he just gives
an order of magnitude comparison with his wind force expression.\cite{sham}
This has led him to the conclusion that the screening is negligible.
We find a screening of 5-25\%, which is in agreement with
Sorbello's results.

%The units used in the paper can be traced only through the absence
%of $\hbar$ in the exponentials, because $\hbar=1$.
Rydberg atomic units are used, in which the
energy is in Rydberg, the distance is in Bohr (1 Bohr $\approx$
0.5 \AA), $\hbar=1$, and the electronic mass is $\frac{1}{2}$.

\section*{Basics of linear-response theory}

We first give the standard linear-response expression for the driving force
on an ion at position ${\bf R}_i$,\cite{lod90}
\begin{equation}
\label{dforce}
{\bf F}=Z_i e {\bf E}-ie E_\nu\int_{0}^{\infty}
{\rm d}t e^{- at}Tr\Big\{ \rho (H)
\bigg[{\bf F}_{\rm op} (t) , \sum_j r_j^\nu \bigg] \Big\},
\end{equation}
in which the cartesian label $\nu$ runs from 1 to 3. The infinitisimally
positive number $a$ represents the adiabatical switch-on of the
electric field, the operator
$\rho (H)$ is the grandcanonical density depending on the system
Hamiltonian $H$, the force operator
${\bf F}_{\rm op}$ stands for
\begin{equation}
\label{FopdVdR}
{\bf F}_{\rm op}\equiv -\nabla_{{\bf R}_i} V=-\sum_j
\nabla_{{\bf R}_i}v({\bf r}_j-{\bf R}_i)\equiv\sum_j{\bf f}_j^i,
\end{equation}
and its time dependence refers to the Heisenberg representation
\begin{equation}
\label{FHeisen}
{\bf F}_{\rm op} (t)\equiv e^{iHt}{\bf F}_{\rm op}e^{-iHt}.
\end{equation}
The expression published by Kumar and Sorbello \cite{kumar}
follows simply after a partial integration
in Eq. (\ref{dforce}) with respect to the time.

All important studies of the driving force have been done for
the electron-impurity system, the Hamiltonian $H$ of which can
be written as a sum of single particle Hamiltonians $h$, so
\begin{equation}
\label{Hh}
H = \sum_j h^{j}~~~~~~~{\rm with}~~~~~~~h=h_0+v=h_0+\sum_\alpha v^\alpha,
\end{equation}
in which the summation in the electron-impurity potential runs over
the positions ${\bf R}_\alpha$ of the impurities.
This allows for a reduction of the many body expression (\ref{dforce})
to the following single particle expression,
\begin{equation}
\label{Windred}
{\bf F}=Z_i e {\bf E}-ie E_\nu \int_{0}^{\infty}
{\rm d}t~e^{- at}~tr\Big\{n(h)[{\bf f}^i (t),r^\nu]\Big\},
\end{equation}
where $n(h)$ is the Fermi-Dirac distribution function in operator form
\begin{equation}
\label{nhop}
n(h) = \frac{1}{e^{\beta(h - \epsilon_{\rm F})} + 1}.
\end{equation}

\section*{The treatment of Bosvieux and Friedel (BF)}

We start to follow BF's treatment, by writing down their system
Hamiltonian
\begin{equation}
\label{sysHBF}
H=-\sum_{\mu =1}^N\frac{\Delta_\mu}{2m_\mu}+V({\bf r}_1,\ldots{\bf r}_\mu,
\ldots{\bf r}_N).
\end{equation}
In applying a small perturbation denoted as $\delta V$ they write down
the perturbed wave function $\Psi$. In their appendix BF employ the idea of
switching on the field adiabatically. For the sake of clarity we give
the explicit form of the perturbing potential due to an applied field,
\begin{equation}
\label{deltaVt}
\delta V(t)=
e {\bf E} e^{at}\cdot \Big(\sum_{j} {\bf r}_{j}
-\sum_{\alpha}Z_{\alpha}{\bf R}_{\alpha}\Big)\equiv
\delta Ve^{at}\equiv\big(\delta V({\bf r})+\delta V({\bf R})\big)e^{at},
\end{equation}
which is zero in the limit $t\rightarrow -\infty$. The
position dependent potentials have been defined for later use.
BF's force expression reads as
\begin{equation}
\label{ForceBF0}
\vec{\phi}=-<\Psi|{\nabla}_i(V+\delta V)|\Psi>,
\end{equation}
in which $|\Psi>$ represents the state of the system as it develops
from its unperturbed ground state $|\psi_0>$ due to the perturbation $\delta V$.
In order to find $|\Psi>$ we solve the time dependent Schr\"{o}dinger equation
\begin{equation}
\label{SchroedBF}
i\frac{\partial\Psi(t)}{\partial t}={\cal H}(t)\Psi(t)
\end{equation}
for the total Hamiltonian
\begin{equation}
\label{HtotBF}
{\cal H}(t)=H+\delta V(t)
\end{equation}
by working in the interaction
representation for $\Psi(t)$, defined as
\begin{equation}
\label{PsiDirac}
\Psi_I(t)\equiv e^{iHt}\Psi(t).
\end{equation}
The equation for $\Psi_I(t)$ becomes
\begin{equation}
\label{EqPsiDirac}
i\frac{\partial\Psi_I(t)}{\partial t}=e^{iHt}(-H+i\frac{\partial}{\partial t})
\Psi(t)=e^{iHt}\delta V(t)e^{-iHt}\Psi_I(t).
\end{equation}
After integrating this equation one finds for $\Psi(t)$ linearly in $\delta V$
\begin{equation}
\label{Psiself}
\Psi(t)=-ie^{-iHt}\int_{-\infty}^t dt' e^{iHt'}\delta V(t')e^{-iHt'}
\Psi_I(-\infty)+e^{-iHt}\Psi_I(-\infty).
\end{equation}
With $\delta V(t)=\delta V e^{at}$, applying the
substitution $t-t'\equiv s$, and considering an arbitrary time in the
present, so $t=0$, this becomes
\begin{equation}
\label{Psiself3}
\Psi(0)\equiv\Psi=-i\int_0^\infty dt e^{-(iH+a)t}\delta V e^{iHt}
\Psi_I(-\infty)+\Psi_I(-\infty).
\end{equation}
If one calculates matrix elements with this $|\Psi>$ the factor
$e^{-iE_0\infty}$ in the state $|\Psi_I(-\infty)>$ drops out so that
just the ground state $|\psi_0>$ remains. BF's Eq. (I.2) for $|\Psi>$
is reproduced by inserting the complete set of eigenstates
of the system Hamiltonian $H$, denoted by $|\psi_n>$. 
The only difference is the presence of
the infinitesimal number $a$. BF have omitted it and just denote that in the
summation over $n$ the value $n=0$, corresponding to the ground state,
has to be excluded. It will become clear that this difference
has a dramatic influence on the final results.

Now we evaluate BF's force expression, Eq. (\ref{ForceBF0}),
linearly in the applied field $\delta V$.
\begin{eqnarray}
\label{ForceBF}
&&\vec{\phi}=-<\psi_0|{\nabla}_i(\delta V)|\psi_0>+
i\int_0^\infty dt~e^{-at}<\psi_0|({\nabla}_iV) e^{-iHt}\delta V e^{iHt}
|\psi_0> +~~{\rm c.c.}\nonumber\\
&&~~~=Z_ie{\bf E}+i\int_0^\infty dt~e^{-at}<\psi_0|
%\Big[e^{iHt}{\nabla}_iV e^{-iHt},\delta V\Big]|\psi_0>,
\Big[({\nabla}_iV),e^{-iHt}\delta V e^{iHt}\Big]|\psi_0>,
\end{eqnarray}
in which we used Eq. (\ref{Psiself3}) for $|\Psi>$ and
Eq. (\ref{deltaVt}) for evaluating the matrix element
$<\psi_0|{\nabla}_i(\delta V)|\psi_0>$.
 
In the further evaluation of this equation we want to follow BF's evaluation
given in their \S I. The integral over the time in the first line of
Eq. (\ref{ForceBF}) is carried out after inserting a complete set $|\psi_n>$
of eigenstates. Further, use is made of the equality
$({\nabla}_iV)=[{\nabla}_i,H]$. This way one obtains
\begin{eqnarray}
\label{ForceBF2}
\vec{\phi}&&=Z_ie{\bf E}-\sum_{n}(E_n-E_0)<\psi_0|\nabla_i|\psi_n>
\frac{<\psi_n|\delta V|\psi_0>}{E_0-E_n+ia}+~{\rm c.c.}\nonumber\\
&&=Z_ie{\bf E}-\sum_{n}<\psi_0|\nabla_i|\psi_n><\psi_n|\delta V|\psi_0>
\frac{E_n-E_0-ia+ia}{E_0-E_n+ia}+~{\rm c.c.}\nonumber\\
&&=Z_ie{\bf E}+<\psi_0|\nabla_i\delta V|\psi_0>-ia\sum_{n}<\psi_0|\nabla_i|\psi_n>
\frac{<\psi_n|\delta V|\psi_0>}{E_0-E_n+ia}+~{\rm c.c.}
\nonumber\\&&=-ia\sum_{n}<\psi_0|\nabla_i|\psi_n>
\frac{<\psi_n|\delta V|\psi_0>}{E_0-E_n+ia}+~{\rm c.c.}
\end{eqnarray}
We want to comment on Eq. (\ref{ForceBF2}) in view of the results of BF. First,
in evaluating the matrix element $<\psi_0|[\nabla_i,H]|\psi_n>$ in the
first line of Eq. (\ref{ForceBF2}) BF create in addition surface integral
terms corresponding to Green's theorem, by that not appreciating the
hermitian property of $H$. These surface integral terms represent flow
of probability out of the system, which is obviously zero for a finite system
and for an isolated metal, as BF admit. But this flow is zero also for a metal
carrying a steady electric current, which is denied in practice by BF.
Secondly, in their treatment the last line
of Eq. (\ref{ForceBF2}) is missing, because BF have put $a=0$ from the
beginning. We want to point out that regarding this last line
a treacherous, hidden trap 
in the formalism is involved, which has shown up earlier in electromigration theory
\cite{sorbelloObj}. This term
seems to be zero because of the proportionality
with the infinitisimal number $a$, but we will show that this
term in fact is a rich one.
 
In rewriting the last line of Eq. (\ref{ForceBF2}) we first write
the energy denominator as coming from a time integral as follows
\begin{eqnarray}
\label{ForceBF3}
\vec{\phi}&&=-a\int_0^\infty dt~e^{-at}\sum_{n}<\psi_0|\nabla_i|\psi_n>
<\psi_n|e^{-iHt}\delta Ve^{iHt}|\psi_0>+~{\rm c.c.}\nonumber\\&&=
-a\int_0^\infty dt~e^{-at}<\psi_0|\Big[\nabla_i,e^{-iHt}\big(\delta V({\bf r})
+\delta V({\bf R})\big)e^{iHt}\Big]|\psi_0>
\end{eqnarray}
Because $H$ commutes with $\delta V({\bf R})$ the second term
reduces to $Z_ie{\bf E}$, and one finds that
\begin{equation}
\label{ForceBF33}
\vec{\phi}=Z_ie{\bf E}-a\int_0^\infty dt~e^{-at}<\psi_0|\Big[
e^{iHt}{\nabla}_ie^{-iHt},\delta V({\bf r})\Big]|\psi_0>.
\end{equation}
After a partial integration with respect to time,
using that $[{\nabla}_i,H]=(\nabla_iV)$, and substituting
$\delta V({\bf r})$ according to Eq. (\ref{deltaVt}), one finds
\begin{equation}
\label{ForceBF4}
\vec{\phi}=Z_ie{\bf E}+i\int_0^\infty dt~e^{-at}<\psi_0|
\Big[e^{iHt}({\nabla}_iV) e^{-iHt},e {\bf E}\cdot\sum_{j}{\bf r}_{j}\Big]|\psi_0>.
\end{equation}
We are back at the second line of Eq. (\ref{ForceBF}), but only
the electron coordinates in
$\delta V$ have survived the operations. The influence of the electric
current on the total force is taken care of by the second term in the
right hand side of Eq. (\ref{ForceBF4}). In retrospection this result follows
rightaway from Eq. (\ref{ForceBF}) as well, but we wanted to go along
with BF's way of evaluation first.

Eq. (\ref{ForceBF4}) is a very interesting
result. It is precisely the zero temperature equivalent of
Eq. (\ref{dforce}). This becomes even more clear if we write down the form
which shows up after the reduction of Eq. (\ref{ForceBF4}) to
single particle states denoted by $|q>$.
\begin{equation}
\label{ForceBFsp}
\vec{\phi}=Z_ie{\bf E}+i\int_0^\infty dt~e^{-at}\sum_q
<q|\Big[e^{iht}({\nabla}_i v) e^{-iht},e{\bf E}\cdot{\bf r}\Big]|q>.
\end{equation}
At $T=0$ the sum over the single particle states has a sharp
cut-off at $\epsilon_q=\epsilon_{\rm F}$. The finite temperature
equivalent of Eq. (\ref{ForceBFsp}) can be written as
\begin{eqnarray}
\label{ForceBFspFT}
\vec{\phi}&&=Z_ie{\bf E}+i\int_0^\infty dt~e^{-at}\sum_q n(\epsilon_q)
<q|\Big[e^{iht}({\nabla}_i v) e^{-iht},e{\bf E}\cdot{\bf r}\Big]|q>
\nonumber\\&&=
Z_ie{\bf E}-i\int_0^\infty dt~e^{-at}tr \bigg\{n(h)\Big[{\bf f}^i(t),
e{\bf E}\cdot{\bf r}\Big]\bigg\}\equiv (Z_{i}+Z_{wind}+Z^{scr})e{\bf E},
\end{eqnarray}
in which the Fermi-Dirac distribution $n(\epsilon)$ has been
inserted, see Eq. (\ref{nhop}).
The force operator ${\bf f}^i$ is defined in Eq. (\ref{FopdVdR}).
Clearly, Eq. (\ref{ForceBFspFT}) is completely equivalent to
Eq. (\ref{Windred}) of the present text. By this electromigration
theory can be considered as being unified. Apparently, BF's starting
formula was correct, but these authors did not recognize its precise contents.
This became even more clear recently.\cite{friedel05} BF used
Eq. (\ref{ForceBF0}) for their result regarding the direct force only.
They treated the wind force separately, applying a semi-classical
standard approach in describing the current carrying electrons
and accounting for the scattering of the electrons by the migrating impurity
quantummechanically.\cite{ziman}
 
The formal Eqs. (\ref{dforce}) or (\ref{ForceBFspFT}), or forms
which have been shown to be equivalent to them, have been
used since Kumar and Sorbello published their linear-response approach to the
electromigration problem \cite{kumar}. The result of all this research
is, that the second term of these equations contains two contributions, as
it has been indicated explicitly in the right hand side of Eq. (\ref{ForceBFspFT}).
One can be identified as the wind force. The other one implies some screening
of the bare direct force $Z_ie{\bf E}$. The wind force expression
has been studied thoroughly and has been applied to calculate
$Z_{wind}$ for both interstitial and substitutional impurities
in many metals.\cite{sor73,DLE97,dekkerJAP} Because the present paper
is devoted to the settlement of the controversy regarding the direct force,
we want to add a calculation of $Z^{scr}$ starting from Sham's
contribution.\cite{sham} But first one more comment on Eq. (\ref{ForceBFspFT})
and its interpretation. BF claim to have proven that for an isolated system
the force on an ion is zero.\cite{friedel05} The present author agrees that this force
is zero, but this fact does not follow from an explicit proof, but from the
knowledge that internally any applied field ${\bf E}$ is screened out by
an electronic surface charge that is built up. This is a general
result from the theory of electromagnetism. That is why then $\vec{\phi}$,
which is proportional to ${\bf E}$, is zero, because ${\bf E}=0$.

\section*{Calculation of $Z^{scr}$}

If one evaluates Eq. (\ref{ForceBFspFT}) to lowest order in the potential $v$
of the ion, for a jellium model of the metal, one finds that\cite{lodphysica}
\begin{equation}
\label{ZscrSham}
Z^{scr}=-\frac{4}{3m}\sum_{kk'}(k^2-{\bf k}\cdot{\bf k}')\frac{|v_{kk'}|^2}
{(\epsilon_{k}-\epsilon_{k'})^2+a^2}\left(\frac{\partial n_k}{\partial \epsilon_k}
-\frac{n_{k}-n_{k'}}{\epsilon_{k}-\epsilon_{k'}}\right),
\end{equation}
in which a $k$ label refers to a plane wave. It has been shown that the
$T\rightarrow 0$ limit is equivalent with Sham's expression.\cite{lodphysica}
Sham gave an order of magnitude estimate by comparing it with
the wind force expression. $Z_{wind}$ is proportional
to the transport relaxation time $\tau$ of the system, while $Z^{scr}$
is proportional to the inverse of an energy, for which the Fermi energy can be chosen.
By that he came out at a ratio of $(\epsilon_{\rm F}\tau)^{-1}\approx 0.01$,
being negligible.

\begin{figure}[htb]
\vspace{8mm}
\epsfig{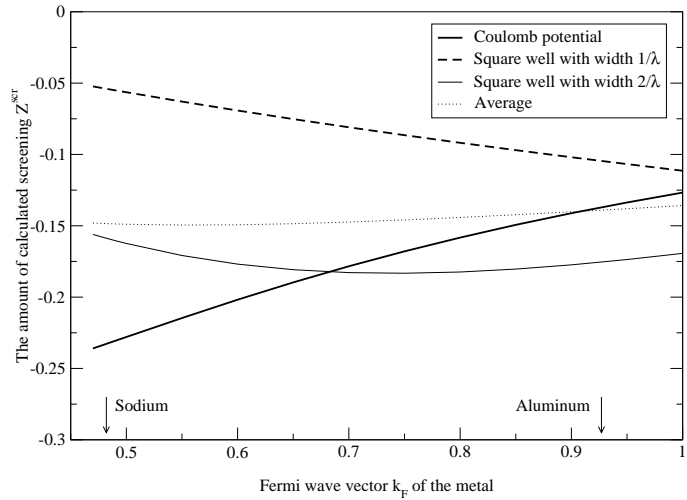}
\caption[]{The amount of screening represented
by $Z^{scr}$ according to Eq. (\ref{ZscrSham3}), with $a=\tau^{-1}$,
for the screened Coulomb potential and for two square well potentials.}
\label{Zscr}
\end{figure}

A numerical evaluation of $Z^{scr}$ becomes possible if one employs
the spherical wave expansion for a plane wave, converts the summations over
the wave vectors to integrals and carries out the angular integrals
over the directions of the wave vectors. After using the relation between
$k^2$ and the energy $\epsilon_k$ one ends up at
\begin{equation}
\label{ZscrSham3}
Z^{scr}=-\frac{4}{3\pi^2m}\int_0^\infty d\epsilon_{k}\int_0^\infty
d\epsilon_{k'}\frac{\frac{\partial n_k}{\partial \epsilon_k}
-\frac{n_{k}-n_{k'}}{\epsilon_{k}-\epsilon_{k'}}}
{(\epsilon_{k}-\epsilon_{k'})^2+a^2}\sum_{\ell}f_\ell(k,k'),
\end{equation}
in which the function $f_\ell(k,k')$ is defined as
\begin{equation}
\label{fellkkp}
f_\ell(k,k')=\epsilon_k\sqrt{\epsilon_{k'}}v_\ell(k',k)
\bigg[(2\ell+1)kv_\ell(k',k)-2(\ell+1)k'v_{\ell+1}(k',k)\bigg],
\end{equation}
containing the information about the ion potential through
\begin{equation}
\label{vellkpk}
v_\ell(k',k)=\int_0^\infty r^2dr~j_\ell(k'r)v(r)j_\ell(kr).
\end{equation}
The integrand has to be treated with care when $\epsilon_{k'}=
\epsilon_k$, because then the denominator attains the
value $a^2$ which would imply 'singular' behaviour. However,
precisely then the numerator becomes zero,
because $\lim_{\epsilon_{k'}\rightarrow\epsilon_k}
(n_{k}-n_{k'})/(\epsilon_{k}-\epsilon_{k'})\rightarrow
\frac{\partial n_k}{\partial \epsilon_k}$. The crucial part
of the integrand lies in the square around the point
$(\epsilon_{k},\epsilon_{k'})=(\epsilon_{\rm F},\epsilon_{\rm F})$.
In studying the $Z^{scr}$ integral it appears that in that square one has to
keep the Fermi-Dirac distribution function in its finite temperature form.
A calculation for $a\ll\tau^{-1}$ has been presented elsewhere.
\cite{condmat} Here we want
to follow Sham more closely. He replaced $a$ by $\tau^{-1}$, by that
accounting for the entire distribution of impurities in the metal and
for the possible presence of other scattering mechanisms.

The result of a numerical evaluation for different ion potentials
is shown in FIG. \ref{Zscr}. In addition to a screened Coulomb potential\cite{ziman}
\begin{equation}
\label{vScreenedC}
v(r)=-\frac{Z_ie^2e^{-\lambda r}}{r},
\end{equation}
with the Thomas-Fermi screening parameter $\lambda$, square well
potentials were employed in the same spirit as Sorbello did.\cite{sor85}
The choice $Z_i=1$ represents a proton in a jellium.
The width of the square well potential was chosen to be equal
to the screening length $1/\lambda$ and twice as large. The corresponding well depth
was limited by the condition that just no bound state could be formed.
The value of $\lambda$ is determined by the Fermi energy. While
Sorbello chose five values for the Fermi energy, typical for metals
ranging from sodium to aluminum, we have done the calculation for
a whole range of Fermi energies. The results are plotted as a function
of the Fermi wave number $k_{\rm F}$. The $k_{\rm F}$ values of
sodium and aluminum are indicated.

Because $\lambda$ increases monotonically with the Fermi energy, the
range of the corresponding screened Coulomb potential decreases
with increasing $k_{\rm F}$, which reduction in strength is seen
clearly in the bold solid curve. We compared $v_{kk}$ for the three potentials
and found a clear decrease for the Coulomb potential
with increasing $k_{\rm F}$, and a rather flat
behaviour for the square well potentials, the one with
$2/\lambda$ being markedly stronger than the one with the smaller width.
The screening to second order in the impurity potential
appears to be not negligible, but on the average
$15\pm10\%$. For comparison we mention, that the result for a very small
$a$ value shows a screening which is at most 2\% larger then the
present result for $a=\tau^{-1}=0.01$.\cite{condmat}
As a guide for the eye we gave the average
of $Z^{scr}$ for the three potentials as a dotted line.
Interestingly, this result does not imply that Sham's conclusion
of a negligible screening is entirely wrong. If fact, he compared
$Z^{scr}$ with $Z_{wind}$. The ratio $Z^{scr}/Z_{wind}$ is small indeed, but this comes
from the large value of $\tau$. It may be clear that a comparison
with $Z_i$ would have been more appropriate.

In conclusion, we have shown that the starting formula of
Bosvieux and Friedel gives the right driving force on an ion in
a metal under the influence of an applied voltage.
Because it always has been recognized as the first time
that a quantummechanical form for that force was written down,
their contribution can still be characterized as a pioneering one.
On the other hand, their prediction of a complete cancellation
of the direct force
has been falsified, because it was based on surface integral
terms, which are zero due to the hermiticity of the
system Hamiltonian. Further, the expression for the magnitude
of the screening due to Sham does not give a negligible screening,
but a screening of about 15\%. This is in agreement with an earlier
result based on another approach.\cite{sor85}

Taking all this together,
the controversy regarding the magnitude of the direct force
can be regarded as being resolved by now, and a unification of the
various descriptions has been achieved.

\acknowledgments

Discussions with Jacques Friedel and an extended recent
correspondence is greatly appreciated.

%\newpage

%\section*{Caption for figures}

%FIG.~\ref{Zscr}. The amount of screening represented
%by $Z^{scr}$ according to Eq. (\ref{ZscrSham3}), for the screened Coulomb
%potential and for two square well potentials.


\begin{thebibliography}{99}

\bibitem{quantum}C.~Bosvieux and J.~Friedel, J. Phys.
Chem. Solids {\bf 23} (1962) 123.
\bibitem{kumar}P.~Kumar and R.~S.~Sorbello, Thin Solid Films {\bf 25} (1975) 25.
\bibitem{sham}L.~J.~Sham, Phys. Rev. B{\bf 12} (1975) 3142.
\bibitem{sor97} For a recent review of the field, see
R. S. Sorbello, in {\it Solid State Physics},
Vol. 51, Eds. H. Ehrenreich and F. Spaepen (Academic Press, San Diego, 1997) 159.
\bibitem{sor85}R. S. Sorbello, Phys. Rev. B {\bf 31} (1985) 798.
\bibitem{rimbey}P.~R.~Rimbey and R.~S.~Sorbello, Phys. Rev. B{\bf 21} (1980) 2150.
\bibitem{lodderKrakow}A. Lodder, Defect and Diff. Forum {\bf 237-240} (2005) 695.
\bibitem{turban}L. Turban, P. Nozi\`{e}res and M. Gerl, J. de Physique
{\bf  37} (1976) 159.
\bibitem{sorbelloObj}R.~S.~Sorbello, Solid State Comm. {\bf 76} (1990) 611.
\bibitem{ishida}H. Ishida, Phys. Rev. B {\bf 51} (1995) 10345.
\bibitem{lod90}A. Lodder, J. Phys. Chem. Solids {\bf 51} (1990) 19.
\bibitem{friedel05}J. Friedel, {\it private communication} (2005).
\bibitem{ziman}J.M. Ziman, {\it Principles of the Theory of Solids},
Cambridge Univ. press, 1992, Chs. 5 and 7.
\bibitem{sor73}R. S. Sorbello,  J. Phys. Chem. Solids
{\bf 34} (1973) 937.
\bibitem{DLE97}J.P. Dekker, A. Lodder and J. van Ek, Phys. Rev. B {\bf 56}
(1997) 12167.
%\bibitem{EL94}J. van Ek and A. Lodder, Defect and Diff. Forum {\bf 115-116} (1994) 1.
\bibitem{dekkerJAP}J.P. Dekker and A. Lodder, J. Appl. Phys. {\bf 84} (1998) 1958.
\bibitem{lodphysica}A.~Lodder, Physica A{\bf 158} (1989) 723.
\bibitem{condmat}A. Lodder, arXiv:cond-mat/0508347 v1 (15 August 2005).
\end{thebibliography}
\end{document}